\documentclass[12pt]{article}

\usepackage{graphicx}
\usepackage{amssymb}

\newcommand{\beq}{\begin{equation}}
\newcommand{\eeq}{\end{equation}}
\newcommand{\beqa}{\begin{eqnarray}}
\newcommand{\eeqa}{\end{eqnarray}}
\newcommand{\beqar}{\begin{eqnarray*}}
\newcommand{\eeqar}{\end{eqnarray*}}

\begin{document}
\thispagestyle{empty}

\vspace{32pt}
\begin{center}
{\textbf{\Large Single Vectorlike Quark Production at the LHC
}}

\vspace{40pt}

R.~Barcel\'o, A.~Carmona, M.~Chala, M.~Masip, J.~Santiago
\vspace{12pt}

\textit{
CAFPE and Departamento de F{\'\i}sica Te\'orica y del Cosmos}\\ 
\textit{Universidad de Granada, E-18071, Granada, Spain}\\
\vspace{16pt}
\texttt{rbarcelo@ugr.es, adrian@ugr.es, miki@ugr.es, 
masip@ugr.es, jsantiago@ugr.es}
\end{center}


\vspace{40pt}

\date{\today}

\begin{abstract}

A gluon resonance $G$ of mass below 1 TeV could be the 
origin of the $t \bar{t}$ forward-backward asymmetry observed 
at the Tevatron provided that new decay modes $G\rightarrow 
q \bar Q$, 
with $q$ a standard quark and $Q$ its massive excitation,
make $G$ broad enough. We consider all the
different cases, with $q$ the top, the bottom or a light quark
and dominant decay modes $Q\to W q'$ or $Q\to Z q$.
We show that current experimental
searches are unable to probe the model, but that minimal departures from 
these analyses can explore a large region of its parameter 
space for the current LHC luminosity. This includes the 
challenging case with the new quarks decaying mostly into light 
quark flavors. In some channels not only the heavy quark but also the massive
gluon can be reconstructed, which would stablish the origin of the
$t\bar{t}$ asymmetry. Similar analyses can be applied to
more general models with new massive gluons and vectorlike quarks.

\end{abstract}


\newpage

\section{Introduction}
With over 10 fb$^{-1}$ of recorded data at the Tevatron and more than
4 fb$^{-1}$ at the Large Hadron Collider (LHC), physics beyond the
standard model (SM) is currently being searched with a very important
degree of detail. Until now no discovery has been
reported by any experimental collaboration, and bounds on many
extensions of the SM rise up to the TeV scale and sometimes higher. 
As an alternative, 
these results may just imply that the experimental signature of the
new physics is peculiar and easy to miss despite being at relatively 
low scales, as preferred by naturalness arguments.
In this article we take this approach to study the $t\bar{t}$
forward-backward (FB) asymmetry at the Tevatron~\cite{AFBTEV}, arguably the most
intriguing departure from the SM predictions. 
We show that it can be
explained by new physics below 1 TeV that could be difficult to 
see in the first round of experiments unless dedicated analyses are
considered. 

Due to the relatively large coupling of the top quark to the
electroweak (EW) symmetry breaking sector, new physics stabilizing the
latter could also appear in top-quark observables. 
This generic argument makes the $2-3~\sigma$ deviation versus
the standard value in the Tevatron asymmetry specially interesting.
Even if it is not
statistically significant at the level of discovery, 
the consistency among different CDF and D0 measurements 
strengthens the case for new physics. 
However, any candidate responsible for the
asymmetry has to be carefully disguised, as its large contribution there
should not translate into any significant departure from the SM in
other related observables. In particular, 
the $t\bar{t}$ total cross section, its 
invariant mass distribution, dijet production, same sign top
production, or the $t\bar{t}$ charge asymmetry at the LHC are
observables where correlated anomalies could be expected 
\cite{generalconstraints}.

In a recent work~\cite{Barcelo:2011vk} 
we have shown that an $s$--channel gluon
resonance $G$ of relatively low mass ($M_G\lesssim 1$ TeV) could
explain the large value of the asymmetry consistently with
all the other observations. (See~\cite{axigluon} for alternative
explanations of the Tevatron asymmetry in terms of massive gluons and
ways to discover them.) 
It should have  small-close to 
axial couplings to the light quarks ($g^q_L\approx -g^q_R$)
together with a large coupling to the right-handed 
top quark, features that are obtained in  Higgsless
models after imposing consistency with EW precission
data~\cite{Barcelo:2011fw}. The key ingredient would be a large
gluon width, $\Gamma_G=(0.5$--$0.7) M_G$,  
provided by new decay modes of type
$G\to Q \bar q , q \bar Q$, where $q$ is a standard quark and $Q$
a massive vectorlike excitation. In composite holographic models
these fields can be understood as Kaluza-Klein modes of 
the standard quarks.
The large gluon width in this framework requires a proper 
treatment of energy-dependent effects. In particular, a Breit-Wigner
with constant width would offer a poor description of the gluon-mediated
amplitude. Instead, when a new channel 
\beq
q\bar q\rightarrow G\rightarrow Q \bar q ,\, q \bar Q\;
\label{gt}
\eeq
opens at $\sqrt{\hat s}= m_Q+m_q$ 
it contributes to  $\Gamma_G(s)$
\beqa
\Gamma_G^{Q q}(\hat s)&=\theta\left[\hat s-(m_q+m_Q)^2\right]\;
 \displaystyle {g^2 \over 12 \pi} {\hat s \over M_G}\, 
\displaystyle \left(1 - {(m_q+m_Q)^2\over \hat s} \right)^{1\over 2} 
\left(1 - {(m_q-m_Q)^2\over \hat s} \right)^{1\over 2}\times
\nonumber \\ 
&
\displaystyle \left[ \left( 1 - {m_q^2+m_Q^2+6 m_q m_Q\over 2 \hat s } -
{(m_Q^2 - m_q^2)^2\over 2 \hat s^2 } \right) \right. g_A^{Qq\,2} +
\nonumber \\
&
\displaystyle \left. \left( 1 - {m_q^2+m_Q^2-6 m_q m_Q\over 2 \hat s } -
{(m_Q^2 - m_q^2)^2\over 2 \hat s^2 } \right)g_V^{Qq\,2}  \right] 
\;, 
\label{GtT}
\eeqa
where $g_{V,A}^{Qq}=(g_R \pm g_L)/2$ are the vector and axial 
coupling of the massive gluon
to $Q$ and $q$, respectively. 
The large width will then reduce all gluon effects
at $\sqrt{\hat s}> m_Q+m_q$ (like a peak in the $t\bar t$ or the dijet
distributions) while leaving unchanged lower energy effects
(namely, the Tevatron FB asymmetry). 

For this mechanism to properly explain the asymmetry we
need $700\mbox{ GeV} \lesssim M_G \lesssim 900\mbox{ GeV}$ and
$400\mbox{ GeV}\lesssim m_Q \lesssim 700\mbox{ GeV}$~\cite{Barcelo:2011vk}.
The low masses of the gluon and the new quarks,
together with the sizable couplings required to generate the large width,
make the production of single new quarks mediated by the massive gluon a
very attractive channel at the LHC. 
In this article we investigate its
potential to probe this scenario. The
signal there will depend strongly on the nature of the vectorlike
quark involved. In section~\ref{benchmark} we classify all the 
possibilities and introduce a benchmark model that provides contributions
in all the different channels. In section~\ref{TB} we study single 
vectorlike quark production 
involving the third generation, and in section~\ref{Q} we discuss
the channels with light flavors. In both cases we show
that current analysis could easily miss the model, whereas specific
searches would very likely reveal the 
mechanism responsible for the Tevatron asymmetry. 
Section~\ref{conclusions} is devoted to our conclusions.

\section{A benchmark model\label{benchmark}}

In this section we introduce a benchmark model that successfully
reproduces the Tevatron FB asymmetry with no conflict with other
experimental tests. It contains simultaneously all 
possible decay channels and, therefore, allows us to perform
a comprehensive study of the stealth gluon scenario. The model 
admits variations where one or several channels are suppressed 
while the others are enhanced in such a way that the total 
gluon width does not change significantly. 
We take $M_G=850$ GeV, although similar setups
can be obtained for gluon masses as low as 700 GeV. 
We fix the couplings to $G$ of the SM
quarks to 
\beqa
g_L^q&=&0.3\,g_s, \quad  g_R^q =g^b_R=-0.3\,g_s, 
\quad
g_R^t=+4\,g_s, \quad 
g_L^t=g_L^b=0,
\label{gencoup}
\eeqa
where $g_s$ is the strong coupling constant. 
For the vectorlike quarks, we will assume the presence
of six fields, corresponding to the excitations of  
$t_R$, $b_R$ and the four light flavors $q_L$ (from now on 
we use $q$ and $Q$ just for the $(u,d,s,c)$ quarks and
their excitations). We fix their masses to
\beqa
M_T &=& 450\;{\rm GeV}\,,\quad M_B = M_Q= 600\;{\rm GeV},
\eeqa
and their flavor-changing couplings to the heavy gluon to
\beqa
g_R^{Tt} &=& 4\,g_s\,,\quad
g_R^{Bb} = 3.5\,g_s\,,\quad
g_L^{Qq} = 3.5\,g_s\,.
\eeqa
These values imply a total width $\Gamma_G\approx 0.7\,M_G$ 
and the 
decay branching fractions 
\begin{eqnarray}
&&{\rm BR}(G\to t\bar{t})\approx 0.2,\qquad\quad
\;\;{\rm BR}(G \to T\bar{t},t\bar{T})\approx 0.24,\nonumber \\
&&{\rm BR}(G\to B\bar{b},b\bar{B})\approx 0.11,\quad
{\rm BR}(G\to Q\bar{q},q\bar{Q})\approx 0.44.
\end{eqnarray}

As we mentioned above, the benchmark model 
just defined has the advantage that all
possible channels are present simultaneously. However, when
studing the possible collider implications of this 
scenario we will 
also consider  the extreme cases where all
but one $G$ decay modes are absent:
\begin{eqnarray}
\mbox{Extreme $T$ model:} 
&& g_R^{Tt} = 7.28\,g_s\,,
\quad g_R^{Bb} = g_L^{Qq} = 0,
\label{ext:T}\\
\mbox{Extreme $B$ model:} 
&& g_R^{Bb} = 9.36\,g_s\,,
\quad g_R^{Tt} = g_L^{Qq} = 0,
\label{ext:B}\\
\mbox{Extreme $Q$ model:} 
&& g_L^{Qq} = 4.68\,g_s\,,
\quad g_R^{Tt}=g_R^{Bb} = 0,
\label{ext:Q}
\end{eqnarray}
and all the other couplings unchanged. In these cases 
the heavy gluon has a $20 \%$ 
branching ratio into $t\bar{t}$ and $80 \%$ into the new
channel. Note that in some of these models the required coupling is
unrealistically large. We just take them as limiting examples to get
clear idea of the LHC reach for these signatures (realistic models 
should lie somewhere in between the benchmark and the extreme cases).

The new heavy quarks will then be produced through
$G$ in the $s$--channel as $Q\bar Q$ pairs or as a
single particle together with a standard quark, $Q\bar q$. Pair
production will also receive the standard QCD contribution (in fact,
due to the axial nature of the $G$ coupling to light quarks, the
interference terms cancel and away from the resonance pair production
is like in the SM).
Single heavy-quark 
production, on the other hand, 
is unsuppressed and opens kinematically at lower
energies ($\sqrt{\hat{s}}=m_q+m_Q\ll 2m_Q$),
appearing as a very promising mechanism 
unexplored in previous literature.
The vectorlike quarks will then decay in a 
model-dependent way, according to their electroweak quantum numbers
and their mixing with the SM quarks. Assuming weak couplings, their
width will be narrow, and a simple scaling allows to go from one
model to another. To be definite we will take the  
branching ratios obtained in the large-mass 
limit of the usual Higgsless models, 
\begin{equation}
\mathrm{BR}(Q\to W q^\prime)
=\frac{2}{3}\;,\quad
\mathrm{BR}(Q\to Z q)
=\frac{1}{3}\;.
\end{equation}
Higgs decays can potentially lead to interesting
signatures~\cite{AguilarSaavedra:2006gw} but we
defer the corresponding analysis to future work.
In this paragraph, we have denoted with $Q$ the six vectorlike quarks. 

\begin{figure}[t]
\begin{center}
\includegraphics[width=.5\linewidth]{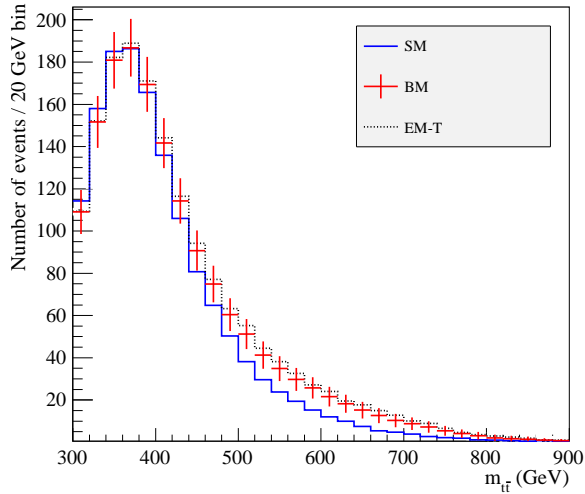}
\caption{\label{mtt_Tevatron} $m_{t\bar{t}}$ distribution at the
Tevatron for 5.3 fb$^{-1}$ in the SM (solid blue), 
the benchmark model (points with error bars) and the extreme $T$ case
(dotted black). We include the contribution from
$T\bar{t},t\bar{T}$ and $B\bar{b},b\bar{B}$ when present.}
\end{center}
\end{figure}

With these assumptions the final states produced in $q\bar q$ collisions 
will be the following (the conjugated processes are not explicitly shown but
are included in our analyses):

\noindent {\it (i)} $W^+W^-b\bar b\,$, from 
\beq
q\bar q\to G\to T \bar t\to  (W^+ b) W^-\bar b
\eeq
and 
\beq
q\bar q\to G\to B \bar b\to (W^-t)\bar b\to (W^-W^+b)\bar b\,.
\eeq

\noindent {\it (ii)} $Zt\bar t\,$, from 
\beq
q\bar q\to G\to T \bar t\to (Zt)\bar t\to (Z W^+ b) W^-\bar b\,.
\eeq

\noindent {\it (iii)} $Zb\bar b\,$, from 
\beq
q\bar q\to G\to B \bar b\to (Zb)\bar b\,.
\eeq

\noindent {\it (iv)} $W\!+\!\mathrm{jets}\,$, from 
\beq
q\bar q\to G\to Q \bar q\to (Wq')\bar q
\eeq

\noindent {\it (v)} $Z\!+\!\mathrm{jets}\,$, from 
\beq
q\bar q\to G\to Q \bar q\to (Zq)\bar q\,.
\eeq

In the next two sections we show that these signals do not introduce
observable anomalies in current LHC analyses, 
but that simple modifications
in the reconstruction of the final state could very likely provide a
signal. The impact of this scenario on top-quark physics at the Tevatron
has been discussed in \cite{Barcelo:2011vk}, where we name it
as the stealth gluon model due to its ability to explain the 
FB asymmetry without introducing anomalies (peaks or tails)
in the $t\bar t$
invariant mass distribution ($m_{t\bar t}$). In particular, it implies 
$A^{t\bar{t}}(m_{t\bar{t}}\leq 450\mbox{ GeV})=0.12$ and
$A^{t\bar{t}}(m_{t\bar{t}}\geq 450\mbox{ GeV})=0.33$, 
values that are 
compatible with the D0 and CDF observations~\cite{AFBTEV}. 
The $m_{t\bar t}$ distribution at the Tevatron
is given in Fig.~\ref{mtt_Tevatron}, where we compare the reconstruction
as $t\bar t$ pairs of all the events giving $W^+W^-b\bar{b}$ in the
benchmark model with 
the SM prediction. In our simulation we have followed the analysis
in~\cite{:2007dia} and have generated the events with MADGRAPH/MADEVENT
v4~\cite{Alwall:2007st} (with the matrix element properly modified to 
include the energy dependence of the width), using
PYTHIA~\cite{pythia} for 
hadronization/showering effects and PGS4~\cite{PGS4} and DELPHES
1.9~\cite{Ovyn:2009tx} for detector 
simulation. We include in the figure the prediction 
in the extreme $T$ model (the prediction in the extreme $B$
model is similar, whereas in the extreme $Q$ model it is below
the benchmark one). The deviations are never larger than
2.5 $\sigma$ (assuming statistical errors only), and below 2 $\sigma$
in all the bins for the benchmark and the extreme-$Q$ models.

\section{Single $T$ and $B$ quark production at the LHC\label{TB}}

\subsection{$W^+ W^- b\bar{b}$ channel}

\begin{figure}[t]
\begin{center}
\includegraphics[width=.6\linewidth]{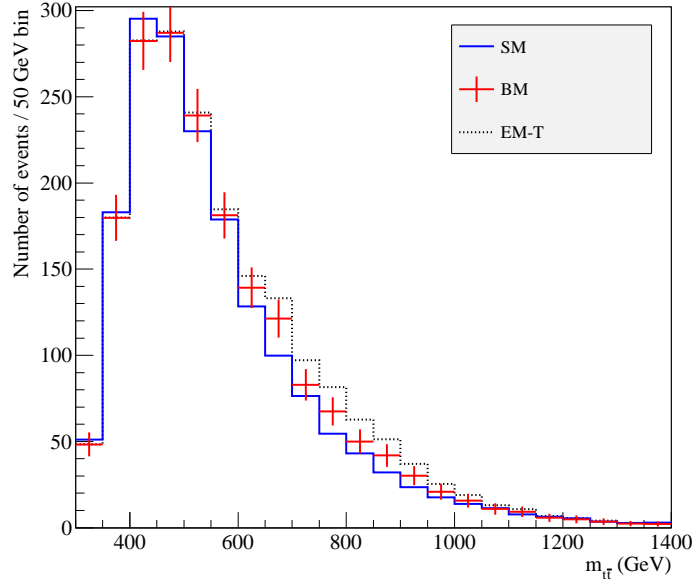}
\caption{\label{LHC_ttbar} $m_{t\bar{t}}$ distribution at the LHC for 
0.2 fb$^{-1}$ in the SM (solid blue), the benchmark model
(points with error bars) and the extreme $T$ model (dotted black). 
We include the contribution from 
$T\bar{t},t\bar{T}$ and $B\bar{b},b\bar{B}$ when present.}
\end{center}
\end{figure}

As described in the previous section, the new processes $q\bar q \to 
T\bar t, B\bar b$ followed by the charged-current decay of the
heavy quark will result in the same
$W^+W^-b\bar{b}$ final state as $t\bar{t}$ production. 
In our model this signal would add to the one from 
top-quark pairs produced
through the massive gluon, and it is then necessary
to check that these processes do not imply any observable
excess in current analyses of $t\bar{t}$ production
or fourth generation $T\bar{T}$ searches. Measurements of the 
$t\bar{t}$ mass invariant distribution at the LHC have been reported
in~\cite{LHCttbar}. 
We have simulated the analyses in the first two works of this reference
and studied the effect of the channels
\begin{equation}
pp\to T\bar{t},\,t\bar{T},\,B\bar{b},\,b\bar{B}
\end{equation}
together with all the contributions to $t\bar{t}$ production. 
We show the result in Fig.~\ref{LHC_ttbar}
(we have followed the second Ref. in~\cite{LHCttbar}
for 0.2 fb$^{-1}$; the third reference uses the dilepton channel
and a larger data set, implying a very similar sensitivity).
In the plot we have assumed a 
$10\%$ uncertainty in the $t\bar{t}$ prediction and allowed a
normalization factor (within this $10\%$) to correctly reproduce the
three bins around the peak at $m_{t\bar{t}}\approx 500$ 
GeV. We show the SM, the benchmark model (with statistical error bars)
and the extreme $T$ model. 
The deviation in the extreme $B$ case
is similar to the one in the extreme $T$ model, whereas 
the extreme $Q$ case is closer than the benchmark to the SM.
The $\approx 20\%$ excess at $m_{t\bar t}=600$--$900$ GeV in
the  extreme $T$ and $B$
models seems in the limit of being probed with the current LHC data.
Increasing the luminosity to 
4 fb$^{-1}$ we find 8 consecutive
bins with differences above 3 $\sigma$ for the DELPHES simulation and
7 consecutive ones for the PGS simulation in the case of the extreme T
model. The benchmark and extreme Q models are not as
clear. For instance, using PGS we find 3 and 2 consecutive bins with departures
larger than 3 $\sigma$ in these cases for a luminosity of 4 fb$^{-1}$ 
(in all our estimates we only include statistical
errors). In summary, in our model one could expect a 10\% excess relative 
to the SM prediction in all the $m_{t\bar t}$ 
bins below 1 TeV. These events are just $t\bar t$ pairs mediated by
the heavy gluon $G$. In addition, the bins between $600$--$900$ GeV
could be increased an extra $15\%$ with 
$T\bar t$ and/or $B\bar b$ events that are reconstructed as $t\bar t$ 
pairs.

\begin{figure}[t]
\begin{center}
\includegraphics[width=.46\linewidth]{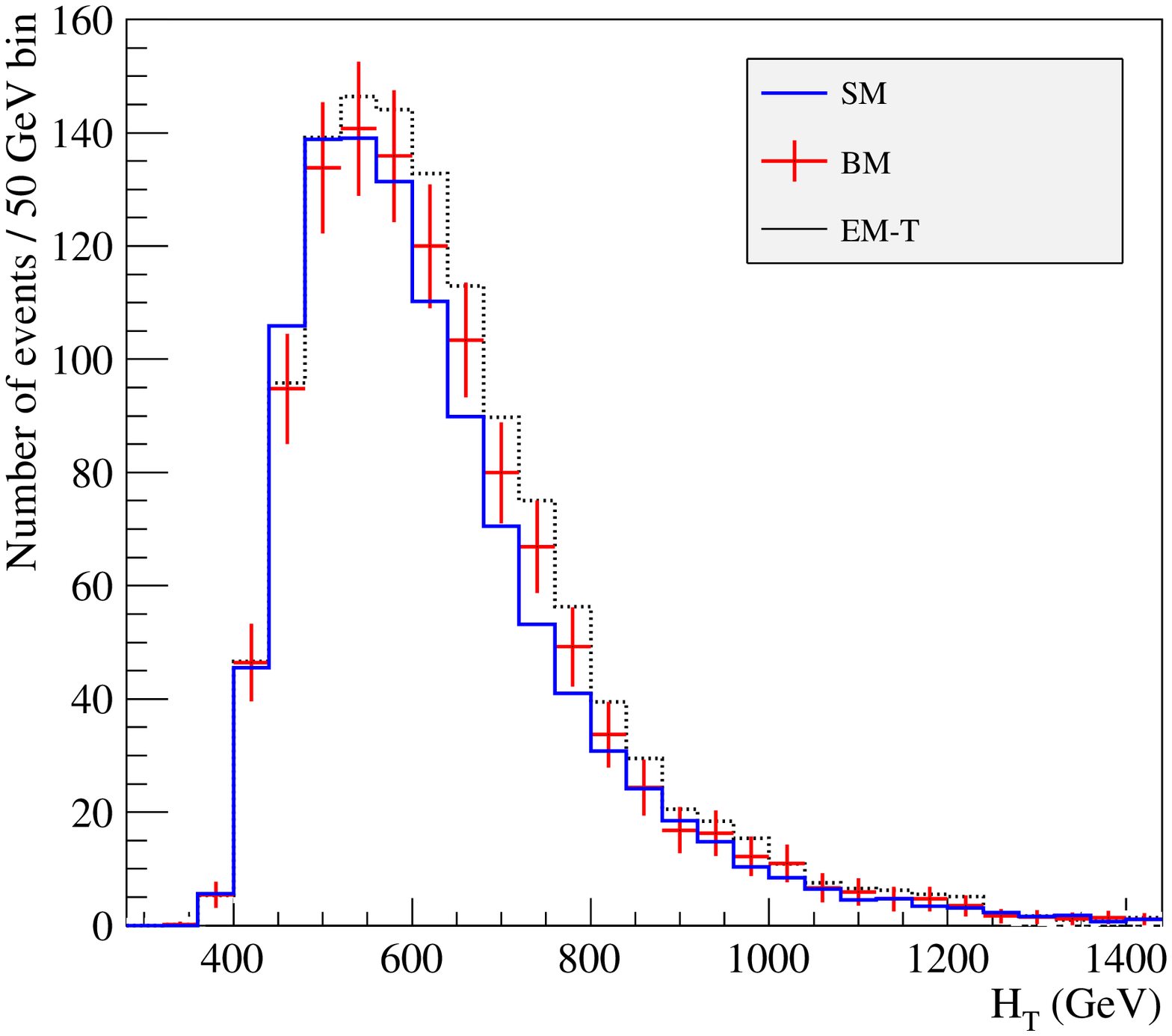}
\includegraphics[width=.46\linewidth]{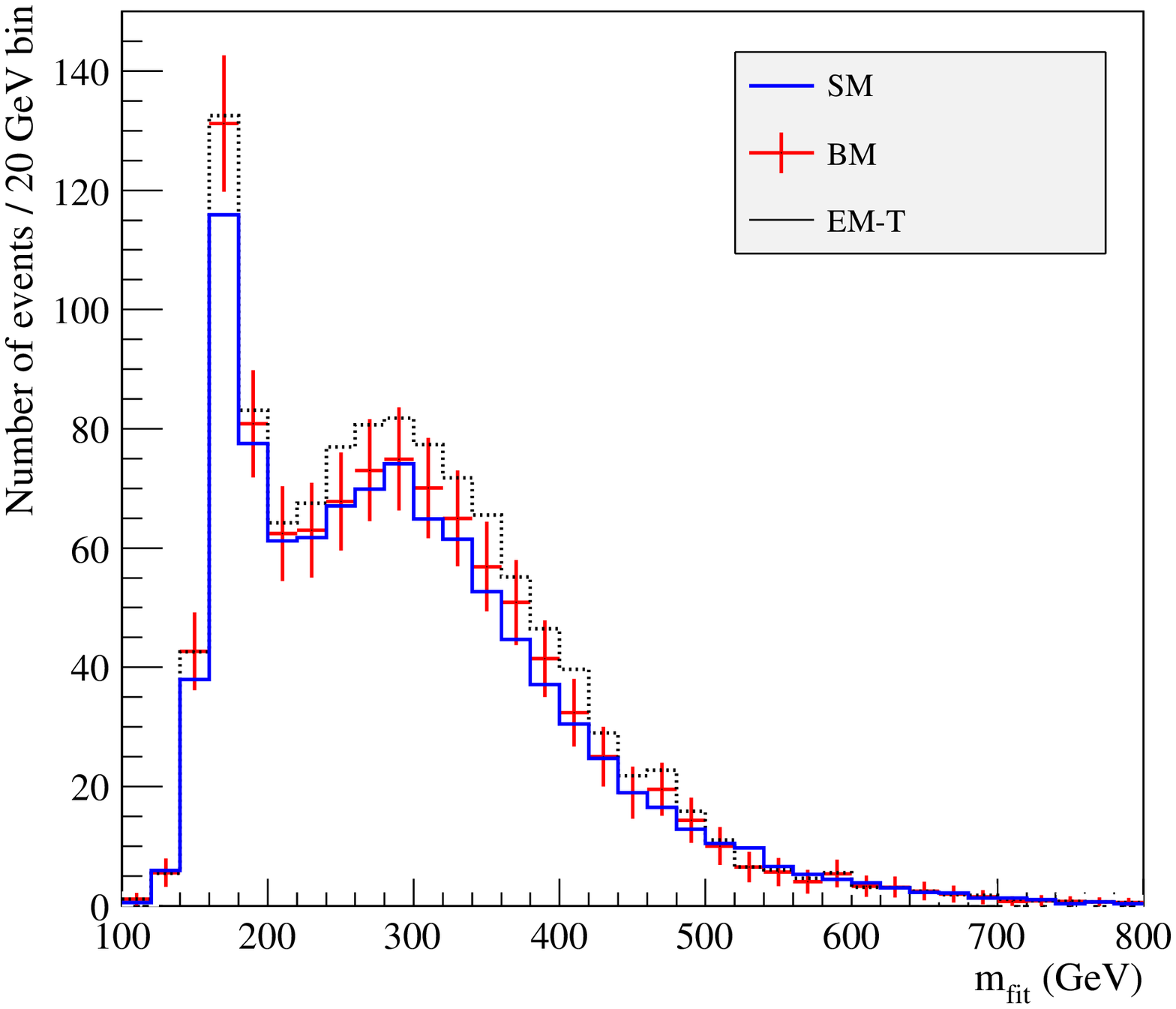}
\caption{\label{LHC_TTbar} $T\bar{T}$ search at the LHC for 0.821
  fb$^{-1}$. Left panel: $H_T$ distribution. Right panel:
  $m_{\mathrm{fit}}$ distribution. In both cases we show the 
predictions in the SM (solid blue), in the benchmark model
(data points with statistical errors) and 
in the extreme $T$ case (dotted black). 
We include the contribution from 
$T\bar{t},t\bar{T}$ and $B\bar{b},b\bar{B}$ when present.}
\end{center}
\end{figure}

Another LHC study sensitive to our model is the search for a fourth 
generation of quarks 
produced as $T\bar{T}$ pairs~\cite{T4thgen}. We have reproduced the
corresponding CMS analysis and plot our results in Fig.~\ref{LHC_TTbar}
for the muon channel with the published luminosity of 0.821 fb$^{-1}$.
Our results are similar to the ones obtained for 
$t\bar{t}$ production. The benchmark and the 
extreme $Q$ models are not visible, whereas the extreme
$T$ and $B$ models are starting to be probed by the data.
We plot in Fig.~3 the SM, the benchmark and the extreme $T$ cases in
solid blue, data points (with error bars), and dotted black, respectively. 
The left panel shows the
$H_T$ distribution (defined in this case as the scalar sum of the $p_T$
of the jets, the charged lepton and the missing $E_T$), 
and the right panel gives the $T$
reconstructed mass in the events generated with 
our model(s) and  with the SM. In both plots the number of
standard events has been normalized by the same factor.
We have also checked that pair production of $T$ quarks give 
in our model a negligible
contribution, compatible with the bound obtained in~\cite{T4thgen}. 
Similarly, the recent search for pair
production of vectorlike $T$ quarks decaying to $Z t$~\cite{TTtoZ} does
not imply any restriction to our model. 

\begin{figure}
\begin{center}
\begin{tabular}{cc}
\includegraphics[width=0.5\linewidth]{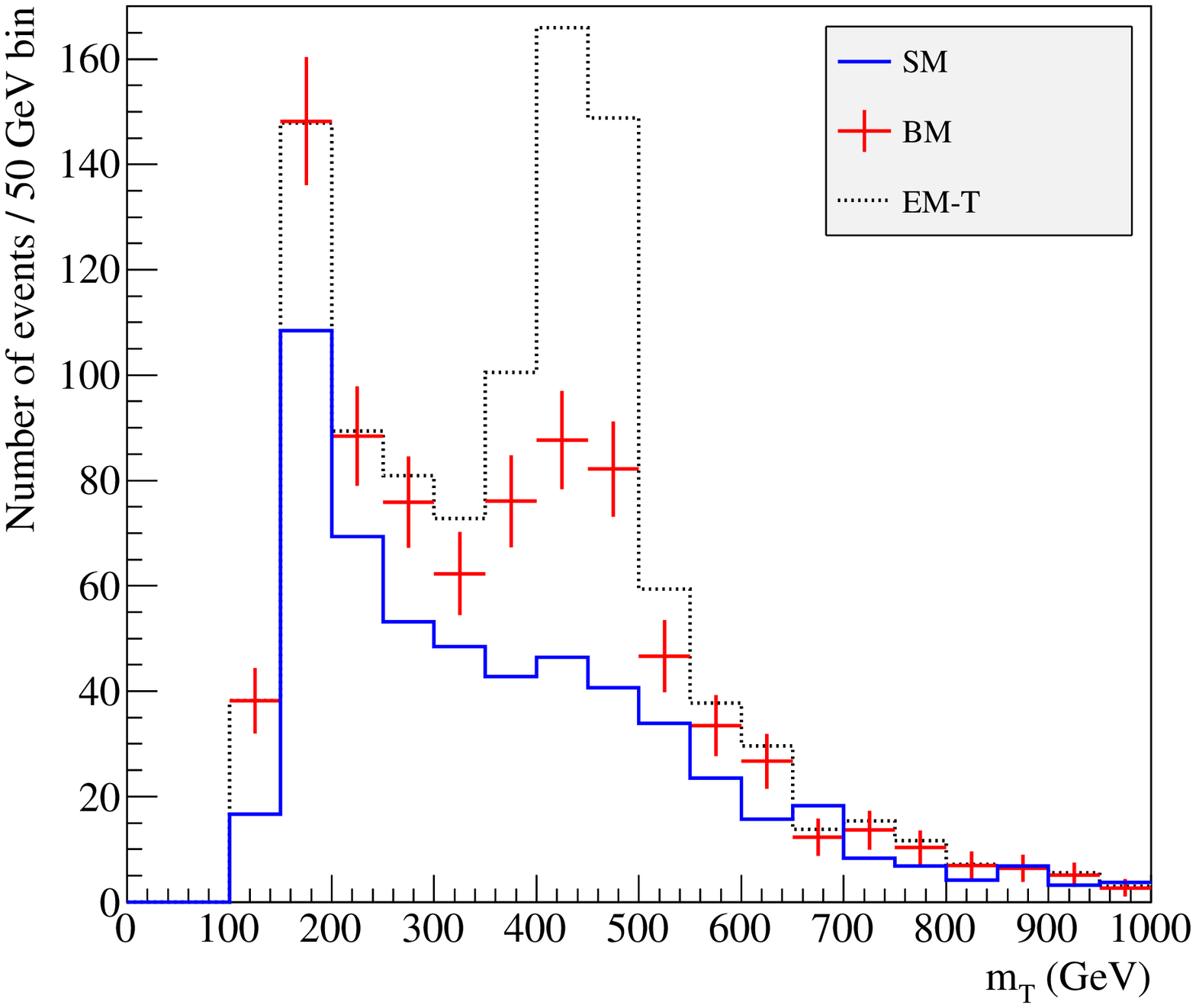} &
\includegraphics[width=0.5\linewidth]{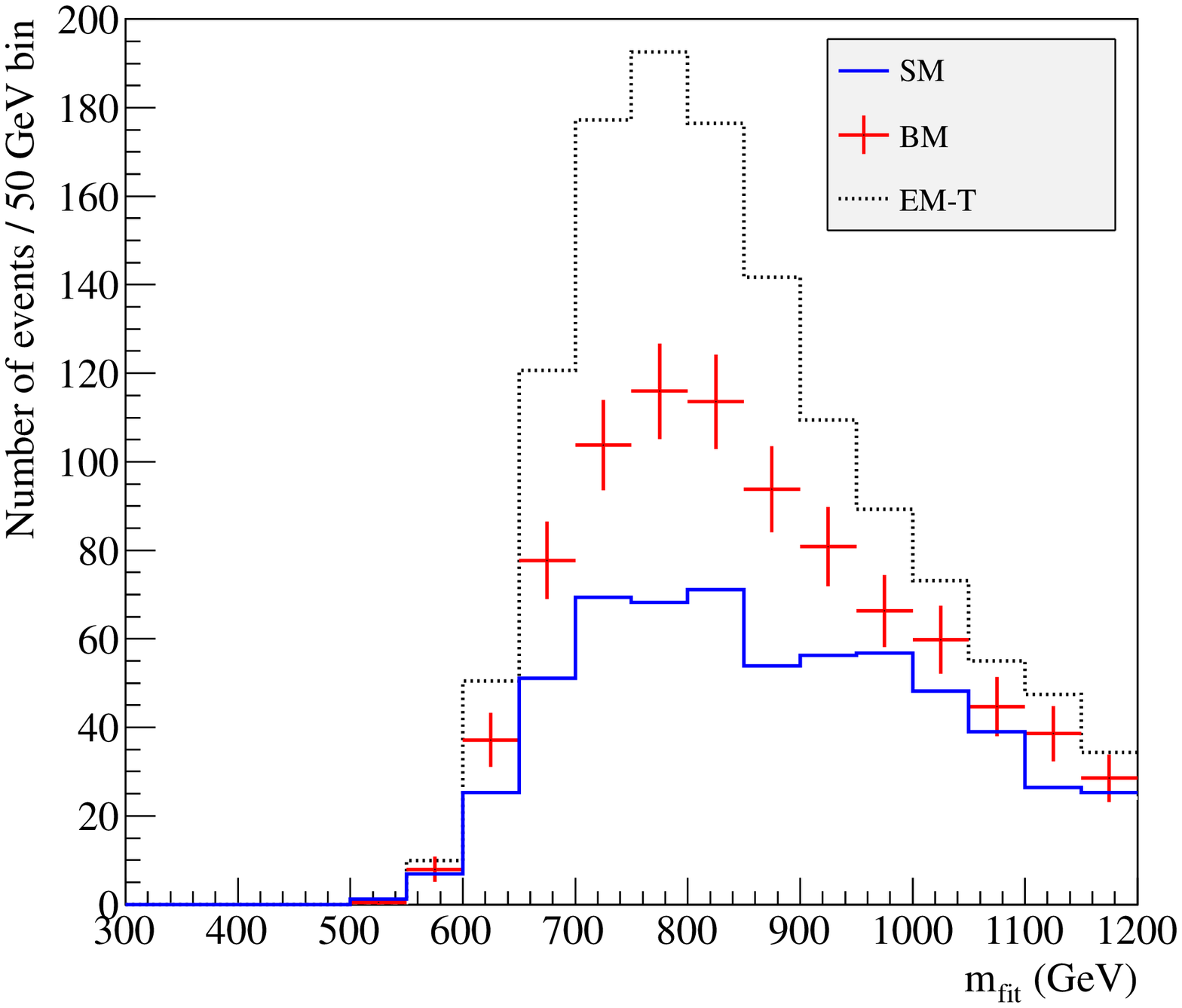}  
\end{tabular}
\end{center}
\caption{\label{T:discovery} Left panel: Reconstruction of $m_T$ at the LHC. 
Right panel: Reconstruction of $m_G$. In both cases we
have normalized the distributions to 4 fb$^{-1}$ data 
and represent the results for the SM
(solid blue), the benchmark model (data points  
with statistical errors) and the extreme $T$ case (dotted black). 
Details of the reconstruction method can be found in the text.
}
\end{figure}

Our results indicate that the model, proposed
to explain the large FB Tevatron asymmetry, is almost invisible in
$t\bar t\to W^+ W^- b\bar{b}$ searches. 
The reason for that is twofold. First, the
large gluon width suppresses the number of $t\bar t$ 
events in the region $m_{t\bar{t}} = 600$--$900$ GeV, while its axial
couplings to the light quarks does the same at lower and higher invariant
masses. Second, $T\bar t$ or $B\bar b$ events 
are reconstructed as $t\bar t$ or $T\bar T$ pairs, resulting into 
a poorer fit and a wider spread.
The key to isolate events of type $T\bar t$ would
be to reconstruct them not like 
two objects with the same mass, 
but like a $t$ quark plus a $T$ quark of arbitrary mass. These
events will only occur at large invariant masses, $m_{T\bar{t}}>m_T+m_t$,
a region already accessible at the LHC with the current
luminosity. Therefore, we 
can use the more stringent cuts used in the $T\bar{T}$ analysis
of~\cite{T4thgen} (we use the muon channel). 
Actually, we will require the hardest jet to have
$p_T\geq 200$ GeV instead of the 120 GeV of that reference.
We will then identify just 
one 173 GeV $t$ quark 
(using a $\chi^2$ similar to the one used in the
first reference of~\cite{LHCttbar} and requiring $\chi^2\leq 10$) 
and will 
plot the mass of the second one in events of invariant
mass above 600 GeV (Fig.~\ref{T:discovery}, left panel) 
for SM and extreme $T$ model simulations. 
We have normalized the plots to the
recorded luminosity of 4 fb$^{-1}$. 
As it is apparent in the plot, 
we find three consecutive bins around $m_T=450$ GeV 
departing more than three sigmas from the SM
prediction even in the benchmark model. 
Counting the total excess $S$ of events versus the standard background
$B$ on the peak (three bins between 350 and
500 GeV) we get
\begin{equation}
\frac{S}{\sqrt{B}}\approx \left\{ 
\begin{array}{ll}
8,\; \quad \mbox{benchmark},\\
21, \quad \mbox{extreme T}.
\end{array}
\right.
\end{equation}
Thus, the extreme $T$ case would 
imply a stunning deviation in this
kind of searches, and even the benchmark model 
could show evidence for new
physics. With the large excess in the extreme $T$ model one can
also try to reconstruct the massive gluon peak. In order to do that, we
remove the total invariant mass 
cut and compute the total invariant mass $m_{T\bar t}$
for the events with a 
reconstructed $T$ mass above 350 GeV. The result is shown in
Fig.~\ref{T:discovery}, right panel. Although the SM and the new
physics model peak in the same region, the factor of $\sim 3(2)$
excess in the extreme B (benchmark) model is quite evident. 

\begin{figure}
\begin{center}
\begin{tabular}{cc}
\includegraphics[width=0.5\linewidth]{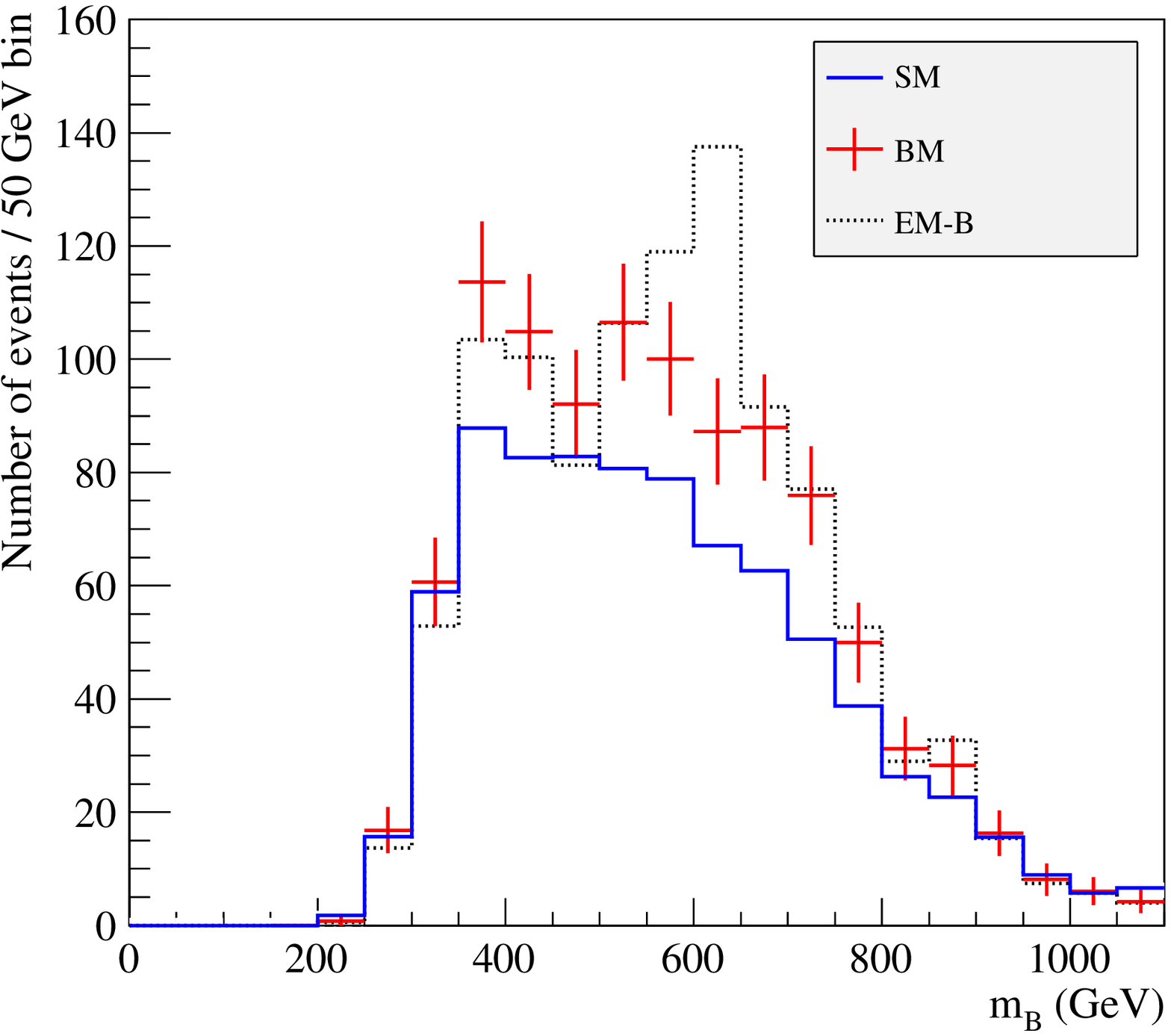} & 
\includegraphics[width=0.5\linewidth]{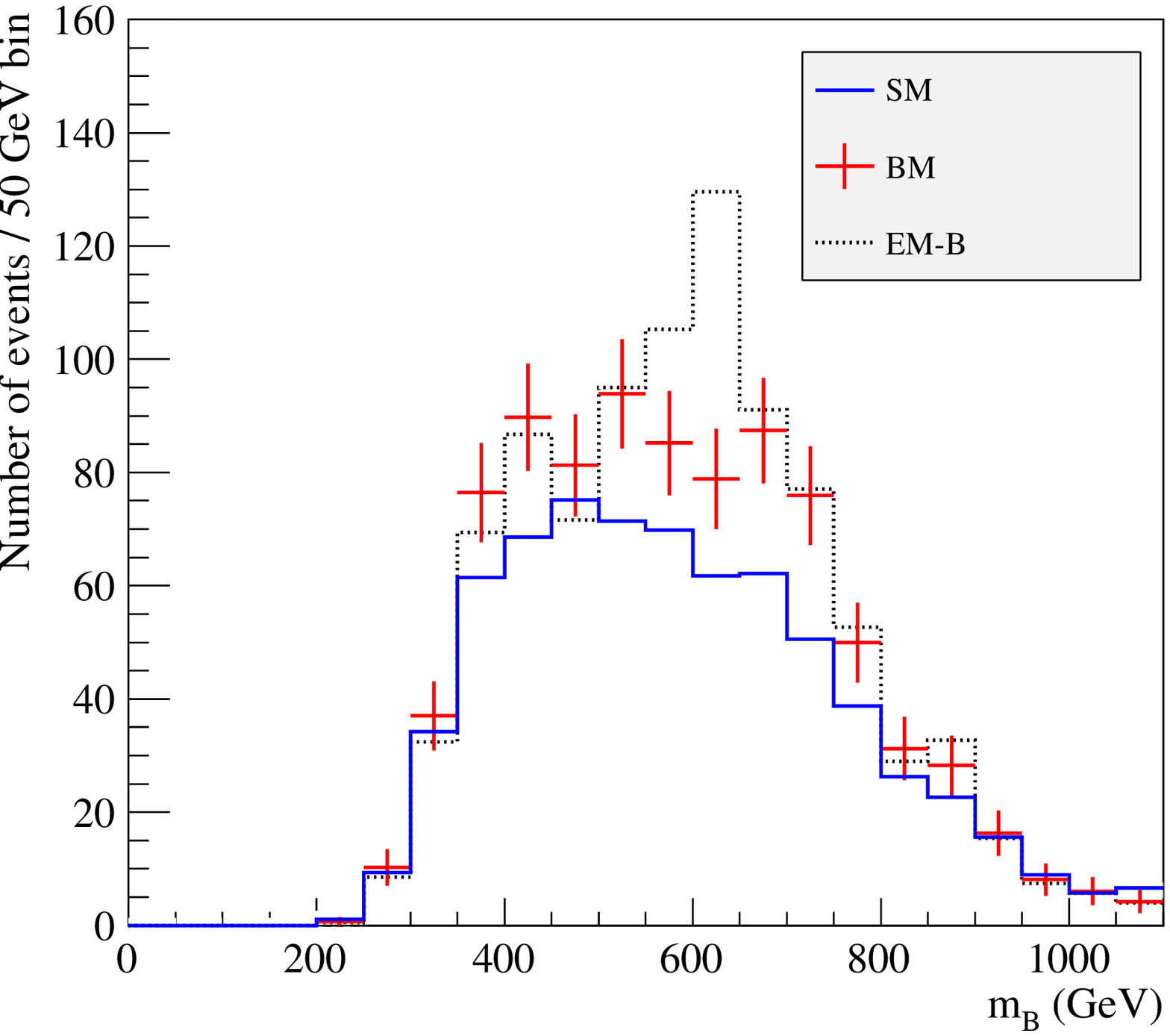} 
\end{tabular}
\end{center}
\caption{\label{B:discovery} Reconstruction of $m_B$ at the LHC for 
4 fb$^{-1}$ in the SM (solid blue), the benchmark model (data points
with errors)
and the $B$ case (dotted black). We consider the cuts $m_{B\bar{b}}>600$ GeV
(left) and $m_{B\bar{b}}>700$ GeV (right).
Details of the reconstruction method can be found in the text.
}
\end{figure}

The $B\bar b,\,b\bar B\to W^+ W^- b\bar{b}$ 
channel is slightly different. Instead of
producing two top-like objects, the heavy bottom decays into a $W$
plus a top that subsequently decays into another $W$ (with opposite
charge) and a $b$. We will still follow the selection procedure in
our previous analysis, with the cuts in~\cite{T4thgen} 
(muon channel) except for the cut on the $p_T$ of the hardest jet,
that is moved from 120 GeV to 200 GeV and a $\chi^2\leq 10$ 
(again we use a similar $\chi^2$ to the one used in the
first reference of~\cite{LHCttbar}) choosing
the best configuration reconstructing a $173$ GeV top quark, 
and will plot the invariant mass of this $t$ quark plus the
extra $W$. The result is shown in Fig.~\ref{B:discovery} for the
benchmark and the extreme $B$ models with two different cuts in the
total invariant mass distribution and 4
fb$^{-1}$. In this case, our reconstruction of the $B$ quark is not as
clear as the one of the $T$ quark, and 
more sophisticated analyses should be used to dig out
the signal from the background. Nevertheless, we will see in the
next section that the extreme $B$ model can be probed much more
efficiently using the neutral decay of the $B$ quark.

\subsection{$Z b\bar{b}$ channel}

\begin{figure}
\begin{center}
\begin{tabular}{cc}
\includegraphics[width=0.5\linewidth]{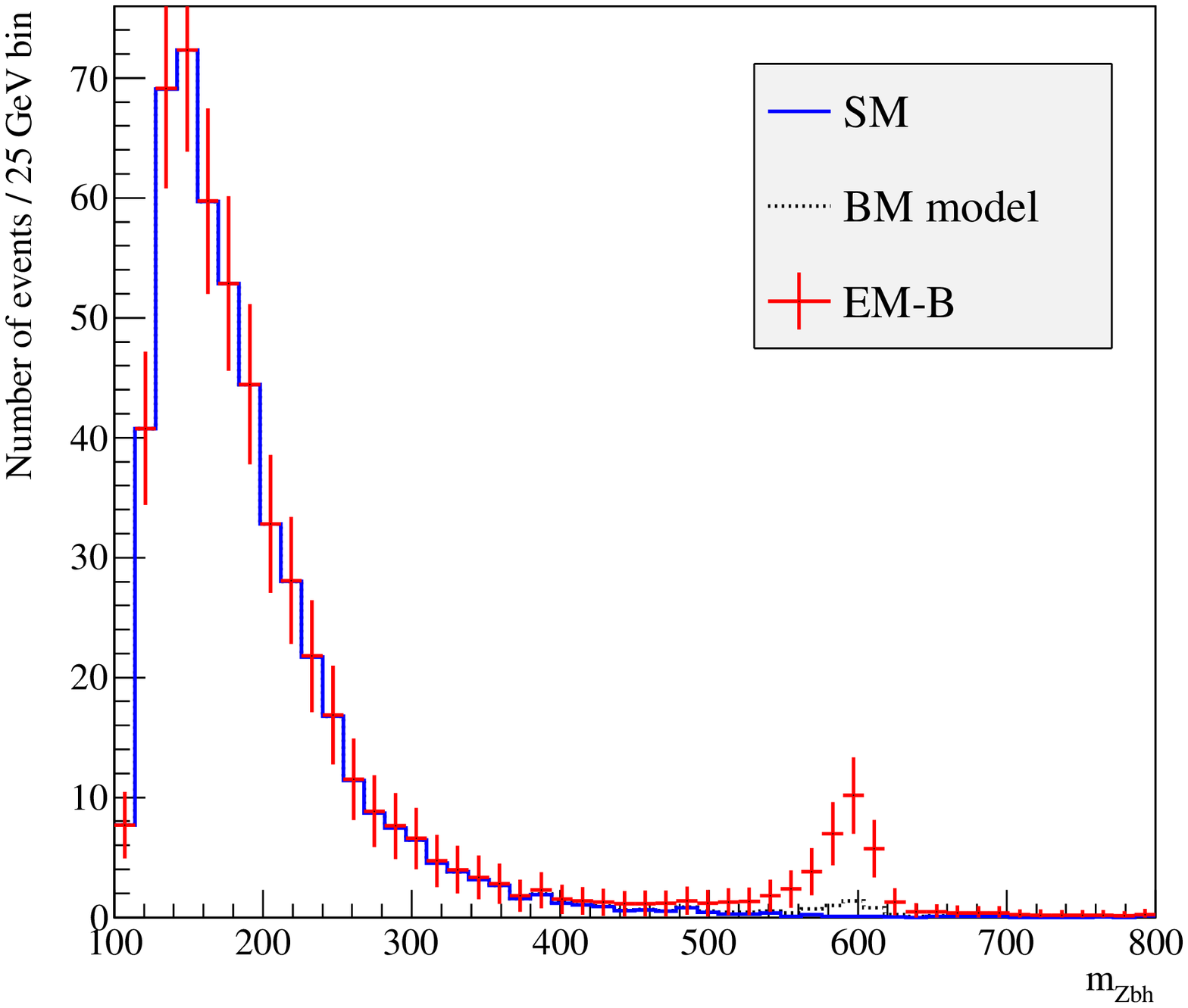}& 
\includegraphics[width=0.5\linewidth]{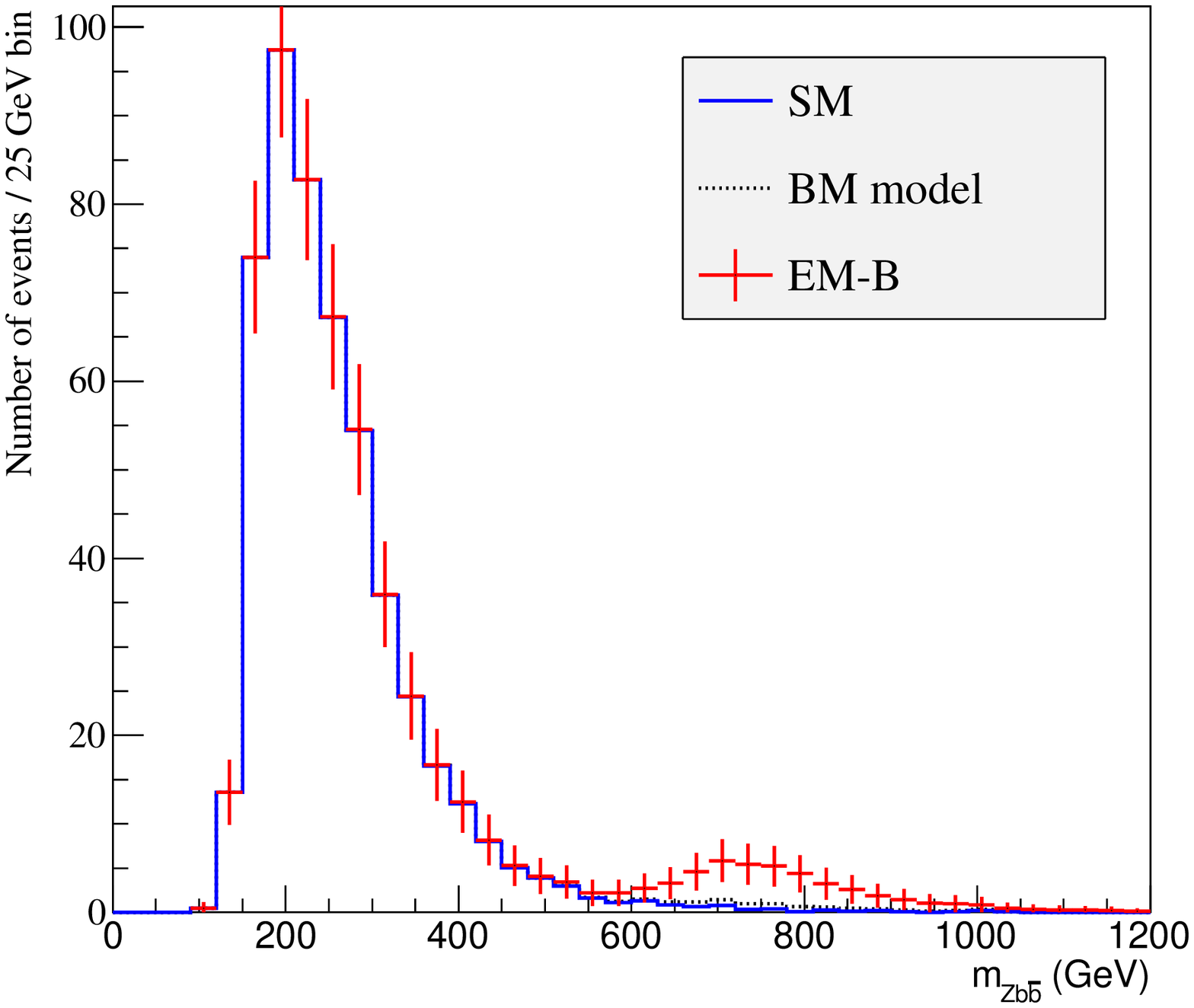}
\end{tabular}
\end{center}
\caption{\label{Zbb:discovery} Left panel: 
reconstruction of $m_{Zb_h}$ at the LHC.
Right panel: reconstruction of $m_{zb\bar{b}}$ to
show the heavy gluon mass.
In both cases we have normalized the distributions to 
4 fb$^{-1}$ of data and have represented the SM with thick solid blue
line, the benchmark model with thin solid red line
and the extreme $B$ case (data points with statistical errors). 
}
\end{figure}

Let us now turn to the neutral decays of the heavy $T$ and $B$
quarks, starting with the $B\bar{b},b\bar{B}$ channel into a
$Z b\bar{b}$ final state. The SM irreducible background to
this process is small ($\sigma(Zb\bar{b})$ with a leptonic $Z$
decay is around $2$ pb), whereas
the background from 
final states with larger cross sections like $Z\!+\!jets$ and 
$t\bar t$ can be reduced with a very simple set of 
cuts.\footnote{We have also checked that our model does not conflict
with current searches of $H\to ZZ\to Z b\bar b$ \cite{Aad:2011ec} or
measurements of $Z+b$ cross-section~\cite{zb:atlas}.}
To isolate the
signal we will require two same-flavor opposite-sign 
leptons with $p_T\geq
25$ GeV and $|m_{l^+l^-}-m_Z|\leq 25$ GeV, and two $b$-tagged jets with
$p_T\geq 20$ GeV and $|\eta|\leq 2.8$. We will also impose 
a veto on missing energy
$E_T\leq 40$ GeV, to reduce the $t\bar{t}$ background. With this
selection we compute the invariant mass of the $Z$ and the hardest of
the two $b$-jets (denoted by $b_h$), since the $b$ quark from the
decay of the heavy $B$  
is typically
the hardest one. We plot the result in Fig.~\ref{Zbb:discovery}. In
the left panel we show the $m_{Zb_h}$ invariant mass distribution in
the SM, the benchmark model and the extreme $B$ case. It is clear that
the distributions in the SM and the new model peak in very different
regions. The benchmark model leads to too small a cross
section and would require higher luminosity for discovery. The extreme
$B$ model, however, shows a clear peak with a total number of $\approx
40$ events at $m_{Zb_h} \approx m_B= 600$ GeV, 
versus $\approx 3$ background events,
implying a statistical significance of  
\begin{equation}
\frac{S}{\sqrt{B}}\approx 21,\qquad (Zb\bar{b}\mbox{ for
extreme $B$}).
\end{equation} 
Given the
presence of a distinct peak we can attempt to reconstruct the mass
of the heavy gluon. In the right panel of Fig.~\ref{Zbb:discovery} we
show the total invariant mass of the three objects $Zb\bar{b}$ for the
events passing the cuts. Due to the large width of the heavy gluon
(the kinematical threshold prevents the full width to be apparent
at energies below $\sim 600$ GeV) the number of events peaks
slightly below $M_G=850$ GeV, but the effect is clearly observable. 
The approximate statistical significance of the excess above
$600$ GeV is 
\begin{equation}
\frac{S}{\sqrt{B}}\approx \frac{38}{\sqrt{5}}=17,
\qquad (\mbox{$M_G$ peak in }Zb\bar{b}\mbox{ for
  extreme B}).
\end{equation} 

The $Zb\bar{b}$ channel appears then as 
very promising even with the very simple cuts that we have used. 
In the extreme case the reconstruction of
the $B$ quark and of the massive gluon at the  4 fb$^{-1}$ LHC could
be correlated with the $t\bar{t}$ anomalies discussed in Section 3.1, 
disentangling the origin of the Tevatron FB asymmetry. 

\subsection{$Z t \bar{t}$ channel}

\begin{figure}
\begin{center}
\includegraphics[width=0.5\linewidth]{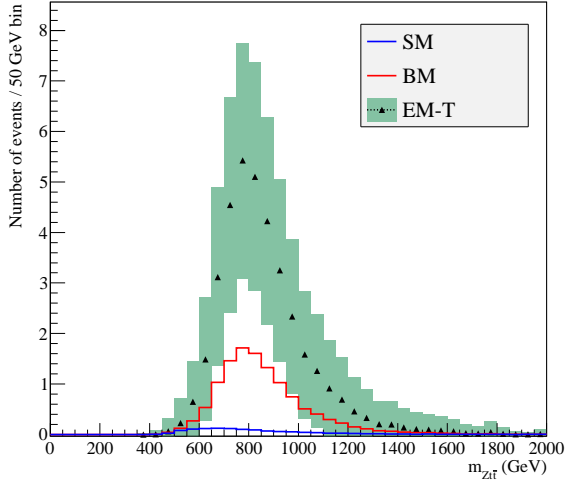} 
\end{center}
\caption{\label{Ztt:discovery} Total invariant mass reconstruction for
  the $Zt\bar{t}$ channel in the SM (solid blue),
  benchmark (solid red) and extreme $T$ (data with statistical
  errors shown as a band) models for the $Zt\bar{t}$ analysis
  described in the text for the LHC with
4 fb$^{-1}$.}
\end{figure}

The $Z t\bar{t}$ production channel resulting into a $Z W^+ W^- b \bar{b}$
final state has also a very small SM background, but it 
is harder to reconstruct due to its
large multiplicity. Instead of trying to
reconstruct the $T$ mass, it is simpler to reconstruct the total final
state in the search for the massive gluon. We do that requiring
{\it (i)} three charged leptons with
$p_T\geq 25$ GeV, and at least two of them with the 
same flavor and opposite sign
reconstructing the $Z$ within 25 GeV; {\it (ii)}
at least two $b$--tagged and at
least two non--$b$--tagged jets with $p_T>20$ and $|\eta|< 2.8$. We
reconstruct the neutrino momentum using the on-shellness condition for
a $W$ and take the two hardest jets and $b$-jets if there are
more of them. The result is shown in
Fig.~\ref{Ztt:discovery}. The extreme $T$ model shows a clear peak
with $\approx 36$ events with no expected background events (the
benchmark gives a weaker deviation). 
A more detailed analysis, trying to 
reconstruct both top quarks, would certainly help in the reconstruction of
the heavy $T$ mass. Since the extreme $T$ model would also 
show up in the charged decay channel, a hint on 
the $T$ mass could be used in the reconstruction of this channel. 

\section{Light flavor excitations: $Wq'\bar q$ and $Zq\bar q$ \label{Q}}

\begin{figure}
\begin{center}
\includegraphics[width=0.45\linewidth]{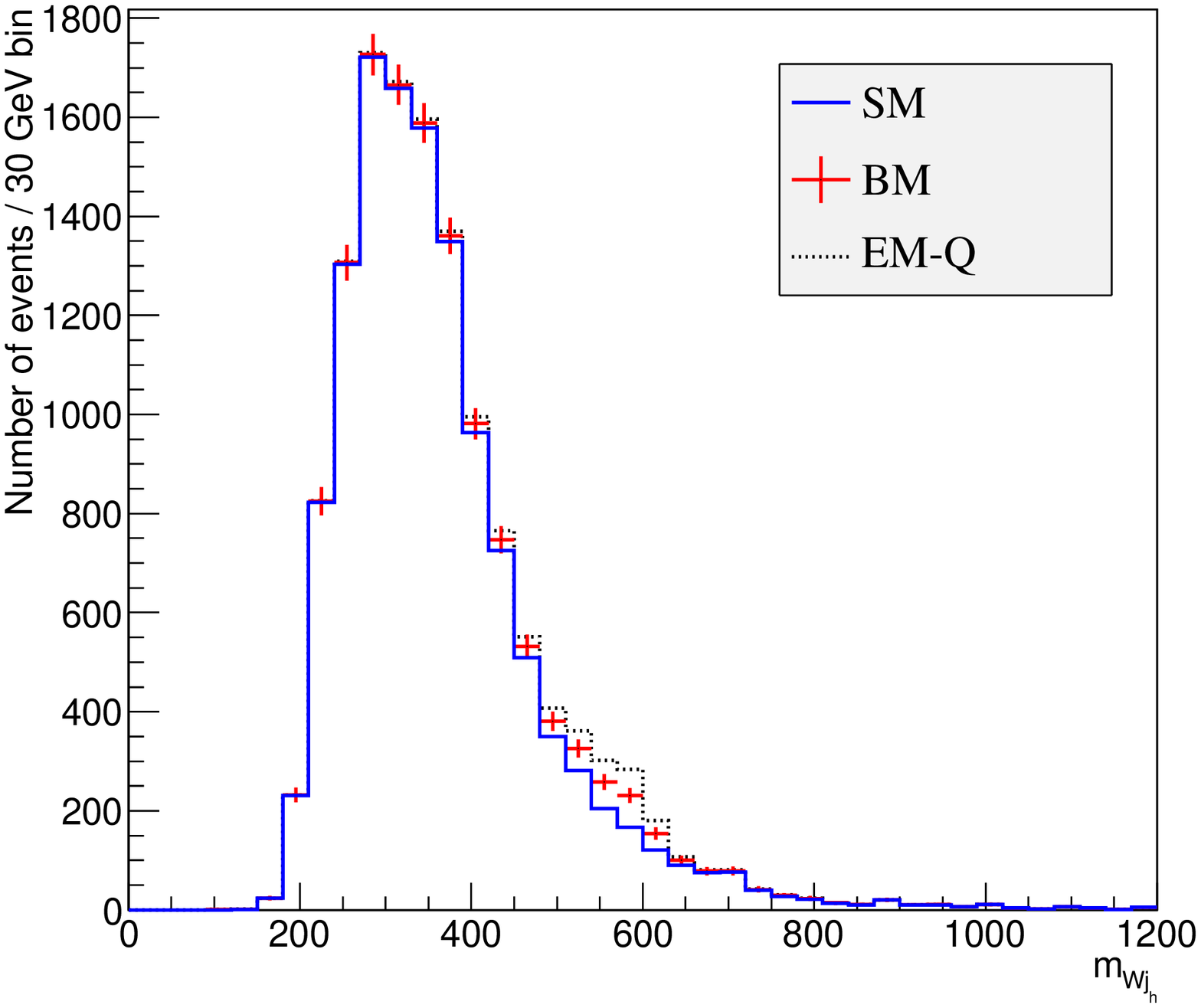} 
\includegraphics[width=0.45\linewidth]{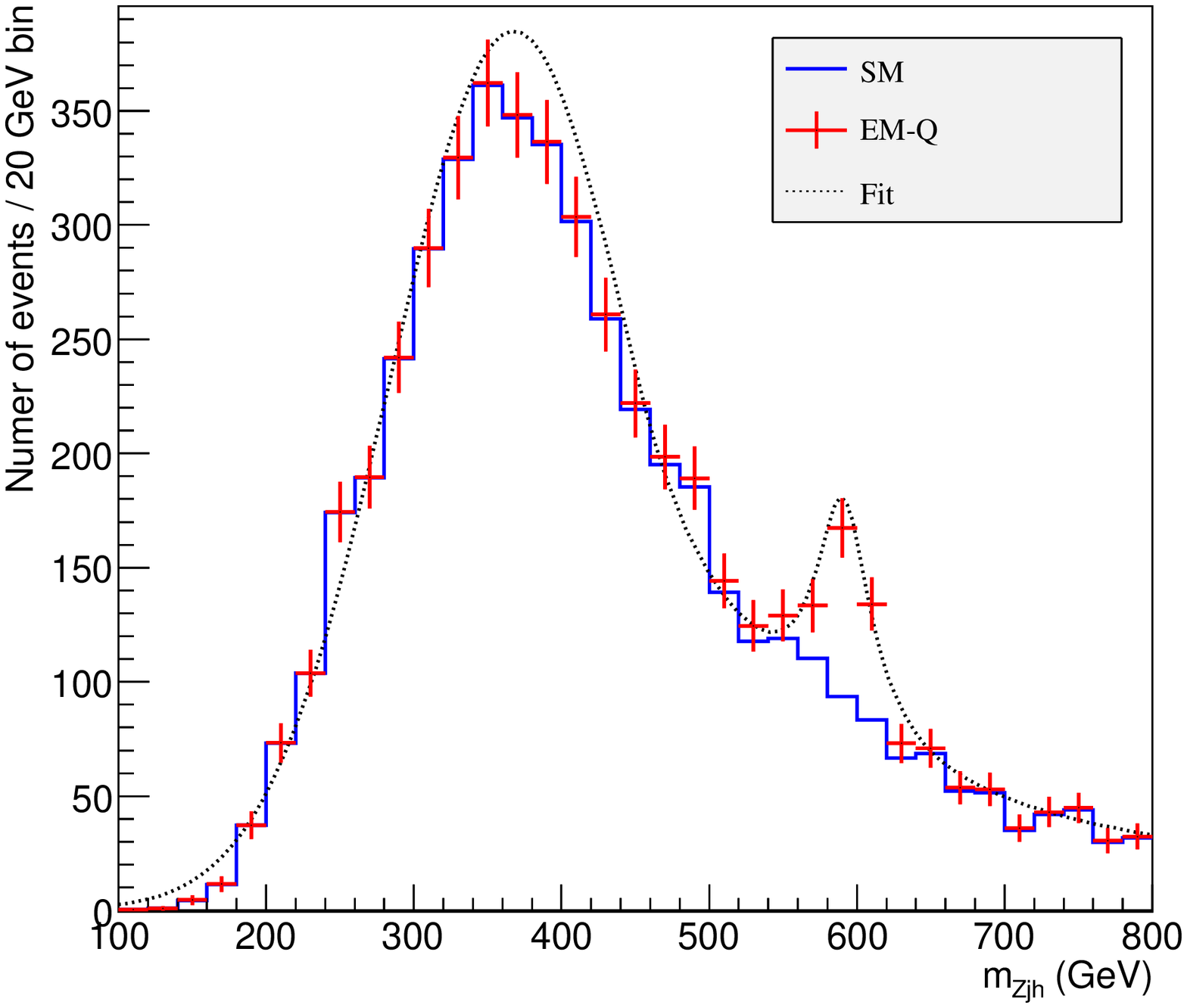} 
\end{center}
\caption{\label{WZj:discovery} Left panel: transverse mass for the $W
  j_h$ system in the $Wjj$ analysis described in the text for the SM
  (solid blue), benchmark model (data points with errors) and extreme
  Q model (dotted black). Right
  panel: Result of the fit of the $m_{Zj_h}$ distribution for the
  $Zjj$ analysis described in 
  the text for the SM (solid blue), extreme Q model (data points with
  statistical errors) and the fit to both distributions (dotted black). 
Both plots are for the 7 TeV LHC with 4 fb$^{-1}$.
}
\end{figure}

We have seen in previous sections that the production of single 
$T$ or $B$ quarks tend to introduce anomalies in current searches 
and could be seen 
if the reconstruction algorithms are slightly modified. 
However, $Q\bar q$ production is less apparent in these searches,
being the best example of stealth new
physics~\cite{Barcelo:2011vk}. 
We discuss in this section the
best strategy to observe the extreme
Q model at the LHC.  
In the benchmark (extreme $Q$) 
model the production of heavy excitations $Q$
of the light flavors has a total cross
section of $2.9$ ($5.4$) pb  at the 7 TeV LHC, resulting
with a 2:1 ratio the final states $Wq^\prime q$ and $Zqq$. 
The SM irreducible background is 17 nb for $W$ plus $\ge 1$ jets 
and 6 nb for $Z$ plus $\ge 1$ jets. 
Therefore, we need to impose stringent cuts to
disentangle our signal from these large backgrounds.
First of all, 
these extra $Q\bar q$ events will only appear at invariant masses 
above $m_Q=600$ GeV, with the maximum at $\approx 700$ GeV.
In addition, the jet from the decay of the heavy
quark, with a $p_T\sim
m_Q/2$, will be typically harder than the second jet.
We should then impose an stringent cut on the hardest jet in order
to reduce the SM backgrounds. In particular, requiring a
hardest jet with $p_T\geq 150$ GeV on top of the cuts defined in
Ref.~\cite{Wjets:ref} reduces the $W\!+\!jets$ background to
manageable levels. We show in
Fig.~\ref{WZj:discovery} (left panel) the transverse mass distribution
of the $W$ and the hardest jet. The signal does not seem significant in the
benchmark model but may be observable in the extreme $Q$ case, 
with 6 bins departing more than 3 standard
deviations from the expected background.

The neutral case is even more promising. Requiring two same-flavor,
opposite-charge leptons with $p_T\geq 25$ GeV that reconstruct the $Z$
mass within 25 GeV, and two or more jets, with $p_T\geq 150$ GeV for
one of them, and computing the invariant mass of the $Z$ and the
hardest jet, we obtain the distribution in
Fig.~\ref{WZj:discovery} (right panel). Although the
benchmark model is still unobservable, there is a clear peak for the
extreme model. We have fitted the signal plus background histogram to
a Crystal Ball plus gaussian shape and obtained an excess of 170 events
over the expected 540 background events in the region of two stardard
deviations around the center of the gaussian. This leads to a
statistical significance of $7\sigma$ and a best fit 
of $m_Q^{\mathrm{fit}}=590$ GeV, very close to the actual heavy
quark mass.
This analysis is interesting as it gives a very clean signal
for a model that is otherwise very difficult to find.

\section{Summary and discussion \label{conclusions}}
The Tevatron $A^{t\bar t}$ anomaly is a strong motivation for
a search of correlated effects from new physics 
at the 7 TeV LHC. An explanation with physics above $\sim 1$ TeV seems
disfavoured by data~\cite{generalconstraints}.
If it is below this energy, 
then a large width could be the key reason why it 
has {\it escaped} detection in the usual observables.
In a recent work we 
proposed that a massive gluon
with new decay modes $G\to Q\bar q$ would be a promising
candidate, and here we have studied in some detail 
the consequences of these processes in current analyses
and possible signals to be searched.

ATLAS and CMS are studying $t\bar t$ and 
$T\bar T$ production. We have shown that
$T\bar t$ production could also be explored just by slightly 
changing the criteria of reconstruction. The channel 
$B\bar b$ provides the same $WWb\bar b$ signal and could
also be studied there.

We have also discussed new channels
that, if analyzed, could reveal single heavy
quark production at the LHC. In particular, 
$Zq\bar q$ where the $Q$ quark is reconstructed with the $Z$ boson
and the highest-energy jet looks promising. Other signals, like
$Zb\bar b$ or $Zt \bar t$, are predicted here and have 
small SM backgrounds.
The study of these channels is well motivated
by holographic models. If the Tevatron
anomaly is due to new quark interactions below the TeV, 
then they should be searched for at the LHC,
since there seems
to be few hideouts for the new physics beneath.
We have focused our study on the region motivated by the Tevatron
asymmetry but our analyses can be also applied to a wider range of
couplings and quark and gluon masses.

\section*{Note added} 
During completion of this work we became aware of Refs.~\cite{roberto}
in which similar ideas to the ones presented here are being
investigated. See also \cite{Dobrescu:2009vz}.

\vspace{0.1cm}

\section*{Acknowledgments}
We would like to thank Nuno Castro, Roberto Contino and James Ferrando
for useful discussions.  
This work has been partially supported by
MICINN of Spain (FPA2006-05294, FPA2010-16802, FPA2010-17915,
Consolider-Ingenio {\bf Multidark} CSD2009-00064 and FPU grant)   
and by Junta de Andaluc\'{\i}a
(FQM 101, FQM 3048 and FQM 6552).


\begin{thebibliography}{99}

\bibitem{AFBTEV}
  V.~M.~Abazov {\it et al.} [ D0 Collaboration ],
  Phys.\ Rev.\ Lett.\  {\bf 100 } (2008)  142002.
  [arXiv:0712.0851 [hep-ex]];
  T.~Aaltonen {\it et al.} [ CDF Collaboration ],
  Phys.\ Rev.\ Lett.\  {\bf 101 } (2008)  202001.
  [arXiv:0806.2472 [hep-ex]];
  T.~Aaltonen {\it et al.} [ CDF Collaboration ],
  [arXiv:1101.0034 [hep-ex]].


\bibitem{generalconstraints}
  M.~I.~Gresham, I.~-W.~Kim, K.~M.~Zurek,
  Phys.\ Rev.\  {\bf D83}, 114027 (2011).
  [arXiv:1103.3501 [hep-ph]];
  C.~Degrande, J.~-M.~Gerard, C.~Grojean, F.~Maltoni, G.~Servant,
  JHEP {\bf 1103}, 125 (2011).
  [arXiv:1010.6304 [hep-ph]];
  C.~Delaunay, O.~Gedalia, Y.~Hochberg, G.~Perez, Y.~Soreq,
    [arXiv:1103.2297 [hep-ph]];
  J.~A.~Aguilar-Saavedra, M.~Perez-Victoria,
  JHEP {\bf 1105 } (2011)  034.
  [arXiv:1103.2765 [hep-ph]];
  [arXiv:1104.1385 [hep-ph]];
    [arXiv:1105.4606 [hep-ph]];
  [arXiv:1107.0841 [hep-ph]];
  C.~Degrande, J.~-M.~Gerard, C.~Grojean, F.~Maltoni, G.~Servant,
  [arXiv:1104.1798 [hep-ph]];
  M.~I.~Gresham, I.~-W.~Kim, K.~M.~Zurek,
  [arXiv:1107.4364 [hep-ph]];
  J.~F.~Kamenik, J.~Shu, J.~Zupan,
  [arXiv:1107.5257 [hep-ph]].



\bibitem{Barcelo:2011vk}
  R.~Barcelo, A.~Carmona, M.~Masip, J.~Santiago,
  [arXiv:1106.4054 [hep-ph]].

\bibitem{axigluon}
  P.~Ferrario, G.~Rodrigo,
  Phys.\ Rev.\  {\bf D78}, 094018 (2008).
  [arXiv:0809.3354 [hep-ph]];
  Phys.\ Rev.\  {\bf D80 } (2009)  051701.
  [arXiv:0906.5541 [hep-ph]];
  A.~Djouadi, G.~Moreau, F.~Richard, R.~K.~Singh,
  C.~Delaunay, O.~Gedalia, S.~J.~Lee, G.~Perez, E.~Ponton,
  Phys.\ Lett.\  {\bf B703}, 486-490 (2011).
  [arXiv:1101.2902 [hep-ph]];
 Phys.\ Rev.\  {\bf D82}, 071702 (2010).
  [arXiv:0906.0604 [hep-ph]];
  P.~H.~Frampton, J.~Shu, K.~Wang,
  Phys.\ Lett.\  {\bf B683}, 294-297 (2010).
  [arXiv:0911.2955 [hep-ph]];
  Q.~-H.~Cao, D.~McKeen, J.~L.~Rosner, G.~Shaughnessy, C.~E.~M.~Wagner,
  Phys.\ Rev.\  {\bf D81 } (2010)  114004.
  [arXiv:1003.3461 [hep-ph]];
  R.~S.~Chivukula, E.~H.~Simmons, C.~-P.~Yuan,
  Phys.\ Rev.\  {\bf D82}, 094009 (2010).
  [arXiv:1007.0260 [hep-ph]];
  G.~Burdman, L.~de Lima, R.~D.~Matheus,
  Phys.\ Rev.\  {\bf D83}, 035012 (2011).
  [arXiv:1011.6380 [hep-ph]].
  E.~Alvarez, L.~Da Rold, A.~Szynkman,
  JHEP {\bf 1105}, 070 (2011).
  [arXiv:1011.6557 [hep-ph]];
 Y.~Bai, J.~L.~Hewett, J.~Kaplan, T.~G.~Rizzo,
  JHEP {\bf 1103 } (2011)  003.
  [arXiv:1101.5203 [hep-ph]];
  A.~Djouadi, G.~Moreau and F.~Richard,
  arXiv:1105.3158 [hep-ph];
  U.~Haisch, S.~Westhoff,
  JHEP {\bf 1108}, 088 (2011).
  [arXiv:1106.0529 [hep-ph]];
  Y.~Bai, Z.~Han,
  [arXiv:1106.5071 [hep-ph]];
  G.~M.~Tavares, M.~Schmaltz,
  Phys.\ Rev.\  {\bf D84}, 054008 (2011).
  [arXiv:1107.0978 [hep-ph]];
  E.~Alvarez, L.~Da Rold, J.~I.~S.~Vietto, A.~Szynkman,
  JHEP {\bf 1109}, 007 (2011).
  [arXiv:1107.1473 [hep-ph]];
  J.~A.~Aguilar-Saavedra, M.~Perez-Victoria,
    [arXiv:1107.2120 [hep-ph]];
  H.~Wang, Y.~-k.~Wang, B.~Xiao, S.~-h.~Zhu,
  [arXiv:1107.5769 [hep-ph]];
  G.~Z.~Krnjaic,
    [arXiv:1109.0648 [hep-ph]];
  J.~A.~Aguilar-Saavedra, A.~Juste, F.~Rubbo,
    [arXiv:1109.3710 [hep-ph]];
  A.~Falkowski, G.~Perez, M.~Schmaltz,
  [arXiv:1110.3796 [hep-ph]].


\bibitem{Barcelo:2011fw}
  R.~Barcelo, A.~Carmona, M.~Masip, J.~Santiago,
  Phys.\ Rev.\  {\bf D84 } (2011)  014024.
  [arXiv:1105.3333 [hep-ph]].

\bibitem{AguilarSaavedra:2006gw}
  J.~A.~Aguilar-Saavedra,
  JHEP {\bf 0612}, 033 (2006).
  [hep-ph/0603200].


\bibitem{:2007dia}
  T.~Aaltonen {\it et al.}  [CDF Collaboration],
  Phys.\ Rev.\  D {\bf 77} (2008) 051102
  [arXiv:0710.5335 [hep-ex]].


\bibitem{Alwall:2007st}
  J.~Alwall {\it et al.},
  JHEP {\bf 0709} (2007) 028
  [arXiv:0706.2334 [hep-ph]].

\bibitem{Mangano:2002ea}
  M.~L.~Mangano, M.~Moretti, F.~Piccinini, R.~Pittau and A.~D.~Polosa,
  JHEP {\bf 0307} (2003) 001
  [hep-ph/0206293].

\bibitem{pythia}
  T.~Sjostrand, S.~Mrenna and P.~Z.~Skands,
  JHEP {\bf 0605} (2006) 026
  [arXiv:hep-ph/0603175].

\bibitem{PGS4} PGS4
http://www.physics.ucdavis.edu/$\sim$
conway/research/software/pgs/pgs4-general.htm 

\bibitem{Ovyn:2009tx}
  S.~Ovyn, X.~Rouby, V.~Lemaitre,
    [arXiv:0903.2225 [hep-ph]].

\bibitem{LHCttbar}
CMS Collaboration, note CMS PAS TOP-10-007;
ATLAS Collaboration, note ATLAS-CONF-2011-087;
note ATLAS-CONF-2011-123.

\bibitem{T4thgen}
CMS Collaboration, note CMS PAS EXO-11-050;
 CMS PAS EXO-11-051;
 CMS PAS EXO-11-054.

\bibitem{TTtoZ}
CMS Collaboration, note CMS PAS EXO-11-005.

\bibitem{Aad:2011ec}
  G.~Aad {\it et al.} [ ATLAS Collaboration ],
    [arXiv:1108.5064 [hep-ex]].

\bibitem{zb:atlas}
ATLAS Collaboration, note CERN-PH-EP-2011-133.

\bibitem{Wjets:ref}
ATLAS Collaboration, note ATLAS-CONF-2011-097.

\bibitem{roberto}
N.~Vignaroli,
  [arXiv:1107.4558 [hep-ph]];
  C.~Bini, R.~Contino and N.~Vignaroli,
  [arXiv:1110.6058 [hep-ph]].

\bibitem{Dobrescu:2009vz}
  B.~A.~Dobrescu, K.~Kong and R.~Mahbubani,
  JHEP {\bf 0906} (2009) 001
  [arXiv:0902.0792 [hep-ph]].
  
\end{thebibliography}
\end{document}